\documentclass[a4paper,11pt]{article}
\pdfoutput=1 

\usepackage{jcappub} 
\usepackage[T1]{fontenc} 
\usepackage[utf8]{inputenc}
\usepackage{graphicx}
\usepackage{epsfig}
\usepackage{rotate}
\usepackage{amsmath}
\usepackage{amssymb}
\usepackage{amsfonts}
\usepackage{bm}
\usepackage{enumerate}
\usepackage{enumitem}
\usepackage{afterpage}
\usepackage{xcolor}
\usepackage{multirow}
\usepackage{array}
\usepackage{natbib, hyperref}
\usepackage[normalem]{ulem}
\usepackage{empheq}

\def\HI{\text{HI}}

\title{A {wide-angle formulation} of foreground filters for HI intensity mapping}

\author{{Rahul} Kothari$^{1,2}$, {Roy} Maartens$^{2,3,4}$}
\affiliation{$^{1}$School of Physical Sciences, Indian Institute of Technology, Mandi, Himachal Pradesh, India 175001 \\
$^2$Department of Physics and Astronomy, University of the Western Cape, Cape Town 7535, South Africa\\
$^3$Institute of Cosmology \& Gravitation, University of Portsmouth, Portsmouth PO1 3FX, UK \\
$^4$National Institute for Theoretical and Computational Sciences (NITheCS), Cape Town 7535, South Africa
}

\emailAdd{quantummechanicskothari@gmail.com}
\emailAdd{roy.maartens@gmail.com}

\abstract{
Neutral hydrogen intensity mapping can in principle deliver rapid and {large}-volume cosmological surveys with exquisitely accurate redshifts that are determined directly from imaging. However, intensity maps suffer from very strong foreground contamination. Future surveys will require efficient data pipelines to remove the foregrounds and reveal the cosmological signal. It is expected that this cleaning will not  remove the signal in substantial parts of the available Fourier space and that significant loss of signal due to imperfect cleaning will be confined to specific regions of Fourier space. This suggests
a strategy which is useful for simplified estimates and rapid computations -- i.e., to apply foreground filters that avoid the regions where loss of signal is significant. {The standard Fourier-space power spectrum and foreground filters use a flat-sky approximation and thus exclude wide-angle correlations.}
We provide a new geometrical formulation of  foreground filters in harmonic space, which naturally includes all wide-angle effects in the power spectrum.   
Foreground filtering leads to a loss of isotropy in 
Fourier space. In {harmonic space} this  produces off-diagonal correlations. We derive analytical expressions for the generalised HI power spectrum and its cross-power with CMB lensing, for both single-dish and interferometer mode surveys. 
We show numerically that the off-diagonal contributions are negligible for
the auto power. In the cross power, there is a non-negligible off-diagonal contribution, but only for a small interval of the largest available scales.
For auto and cross power, the signal loss due to foreground avoidance decreases with increasing multipole (i.e. smaller scales), and the loss in interferometer mode is equal to, or slightly greater than, in single-dish mode.
We find that the cross power in single-dish mode vanishes below a critical multipole, $\ell<\ell_0$. For an SKA-like survey, $\ell_0\sim 20-40$ over redshifts $z=1-3$. {This feature is not seen in interferometer mode as the pertinent angular scales are larger than those allowed by the minimum baseline.} 
}

\begin{document}
\maketitle
\flushbottom

\section{\label{sec:intro}Introduction}
We live in an exciting era of precision cosmology where cosmological models can be robustly tested against  data from  cosmic microwave background (CMB) and large-scale structure (LSS) surveys. 
In addition to the 
next-generation surveys of galaxy number counts, {e.g., {using the}
Dark Energy Spectroscopic Instrument (DESI) \cite{DESI:2016fyo}, Euclid \cite{Euclid:2019clj}, Rubin Observatory Legacy Survey of Space and Time (LSST) \cite{LSSTDarkEnergyScience:2018jkl}, there are {intensity mapping} surveys, e.g. {using the} Square Kilometre Array Observatory Mid-frequency telescope (SKA-Mid) \cite{SKA:2018ckk} and the Hydrogen Intensity and Real-time Analysis eXperiment (HIRAX) \cite{Crichton:2021hlc, Newburgh:2016mwi}, that will target the integrated 21cm spectral line emission of neutral hydrogen (HI), the most abundant element in the Universe.}  
HI intensity mapping (IM) surveys can cover large volumes rapidly as they do not attempt to resolve individual galaxies. In addition, extremely accurate redshifts are obtained directly from imaging, {using the relation between redshift and the frequency at which the image is observed: $1+z = \nu_\mathrm{em}/\nu_\mathrm{obs}$, where $\nu_\mathrm{em}=1.4\,$GHz, with corresponding wavelength $\lambda_{\rm em}\equiv \lambda_{21}=21\,$cm.}

Although HI IM surveys are potentially a very promising cosmological probe, {foreground contamination poses a great challenge and must be removed.}  
{The major source of contamination is Galactic synchrotron emission, due to the acceleration of cosmic ray electrons by the Galactic magnetic field, and $\sim5$ orders of magnitude larger than the 21cm signal. Other sources of contamination include free-free emission (due to acceleration of electrons by ions) and extragalactic bright radio galaxies \cite{Shaw:2014khi, Alonso:2014dhk}.}  

{Sophisticated foreground cleaning techniques have been developed for HI IM (see e.g. \cite{Cunnington:2019lvb,Cunnington:2020mnn,Spinelli:2021emp,Wang:2022bxs, Cunnington:2023jpq}). The main foregrounds are approximately spectrally smooth. Consequently,
a conventional `blind' foreground
clean{ing} will always remove long-wavelength cosmological  modes along the line of sight,  since these modes
will be indistinguishable from the foregrounds.
Radial Fourier modes are $k_\|= \mu k$, where $\mu=\bm n\cdot \hat{\bm k}$ and $\bm n$ is the line-of-sight direction. It follows that foreground cleaning will typically render the modes
$|k_\||< k_{\rm Fg}$ unusable, for an appropriate value of the cut-off $k_{\rm Fg}$. This applies to both  observing modes of HI IM, i.e., single-dish mode (simply add  the signals from all dishes) and interferometer mode (correlate all dish signals).} 

\begin{figure}
\centering
\includegraphics[width=7.5cm]{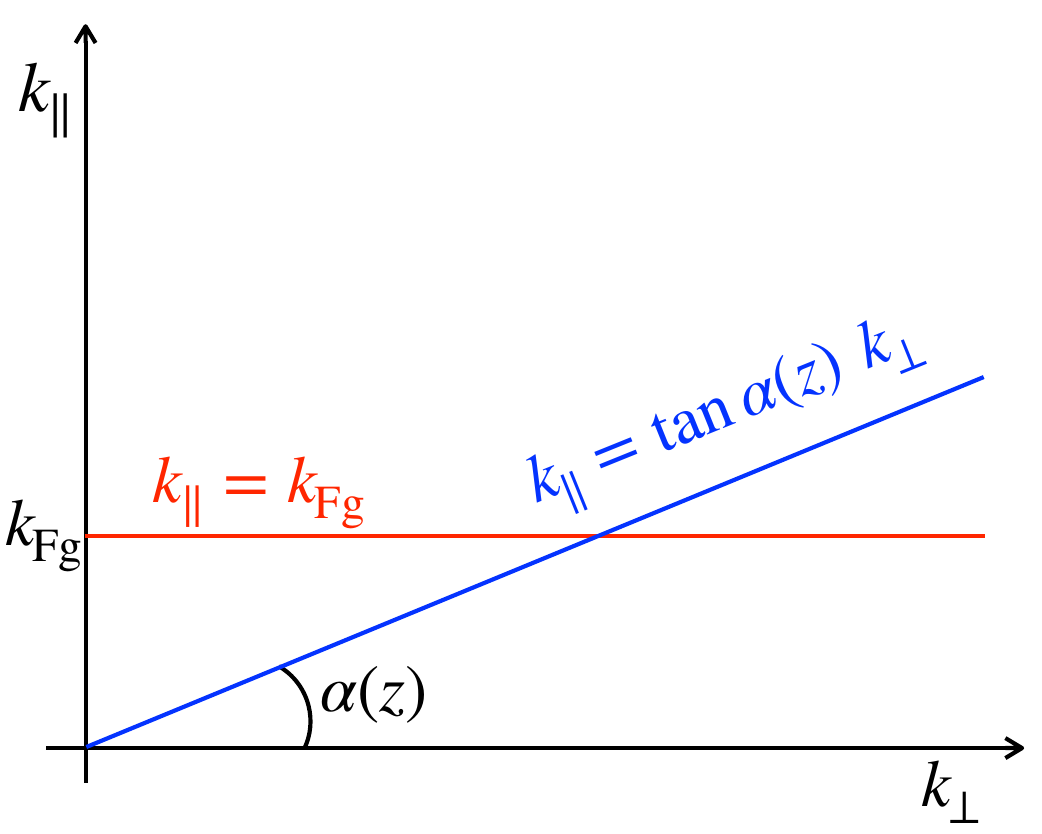}
\caption{\label{fig:FGCond} Schematic of foreground filter limits in Fourier space.}
\end{figure}

{Interferometer-mode surveys suffer from a further problem, known as the `foreground wedge' \cite{CosmicVisions21cm:2018rfq}.
The foreground wedge arises since  interferometer baselines are fixed physical lengths -- which therefore probe different angular scales at different frequencies. As a consequence, intrinsically smooth-spectrum foregrounds can appear to have a more complicated spectra. This affects modes lying in a wedge $|k_\||< (\tan\alpha)\, k_\perp$ in Fourier space, where 
$\alpha$ is described below {(see \autoref{fig:FGCond} and \eqref{eq:wedge})}.
In  principle, the foreground wedge contamination can be reduced by a careful inter-baseline calibration, but this has not yet been achieved \cite{CosmicVisions21cm:2018rfq}.
In summary, a simple foreground filtering technique is to avoid contaminated long-wavelength radial modes, as well as wedge modes in the interferometer case \cite{CosmicVisions21cm:2018rfq, Bull:2014rha, Alonso:2017dgh, Karagiannis:2019jjx, Soares:2020zaq}.}  

{As pointed out in \cite{Byrne:2023rza},  strategies to deal with foregrounds are of two kinds: {(a)} foreground subtraction and {(b)} foreground
avoidance. Foreground subtraction requires modeling the  foreground sources. On the other hand, foreground
avoidance strategies mask the contaminated power spectrum modes. Foreground avoidance, or filtering,
is based on the near-smoothness in frequency of foregrounds -- which means that foregrounds occupy a compact region in Fourier space. 
There are limits to the accuracy of foreground models, which motivates a combination of foreground subtraction and avoidance in data pipelines. In this paper, we focus only on theoretical properties of foreground avoidance.}

Foreground filtering has been described on the basis of the {commonly used  flat-sky (or plane-parallel) approximation in Fourier space, i.e., assuming a global, fixed line-of-sight direction $\bm N$. The HI brightness temperature contrast in Fourier space, $\delta_{\HI}(\bm k,\bm N,z)$ leads to a Fourier power spectrum $P_{\HI}(k,\bm N,z)$, which necessarily neglects wide-angle correlations.} 
In this paper, we find  a new  wide-angle formulation of foreground filtering -- i.e., without assuming a fixed line-of-sight -- and give a geometrical interpretation. We achieve this by working with angular power spectra in harmonic space. {In this approach, the HI temperature contrast is $\delta_{\HI}(\bm n,z)=\Sigma\,a_{\ell m}(z)Y_{\ell m}(\bm n)$, where $\bm n$ is the line of sight to a given pixel at redshift $z$. This leads to an
angular power spectrum that includes all angular correlations.}  As an application, we derive and study the foreground filtered auto-correlation \HI$\times$\HI\ angular power spectrum. 

In addition, we consider the effects of foreground filtering on the angular cross-correlation of \HI\ and CMB lensing $\kappa$. {The lensing of CMB anisotropies contains an imprint of the LSS, projected from $z=0$ to the last scattering surface \cite{Lewis:2006fu}. The combination of CMB lensing with galaxy number counts can improve precision and break parameter degeneracies (see \cite{Chen:2021vba} and references therein). {The situation with HI intensity mapping is different. 
In contrast with galaxies, the correlation between HI intensity mapping and CMB lensing is damped by the effects of foregrounds on  HI intensity.} One might naively expect that unlike auto-correlations \HI$\times$\HI, the cross-correlation \HI $\times\kappa$ will not be affected by HI foregrounds, since these are uncorrelated with $\kappa$. Indeed, many previous studies  neglected foreground effects (e.g. \cite{Sarkar:2009zt, Sarkar:2016uya, Dash:2020yuq, Tanaka:2019nph, Ballardini:2019wxj, Ballardini:2021frp}). However, it is known that the cross-correlation of \HI\ with photometric galaxy surveys is strongly damped -- since HI IM loses long-wavelength radial information while photometric surveys lose short-wavelength radial information \cite{Cunnington:2019lvb, Guandalin:2021sxw, Modi:2021okf}. CMB lensing is a more extreme case of the loss of radial precision on small scales than photometric surveys, since there is no redshift information in the CMB lensing map. {This means that the correlation \HI $\times\kappa$ is 
suppressed even more than \HI $\times\,$photometric galaxy surveys,} on account of the lack of overlap in the large-scale radial modes. {Our findings are consistent with this expectation (see also \cite{Moodley:2023lmu}).}

{In order to compare theory and data, we need window functions to average HI power spectra over a redshift bin $\Delta z$, since  we cannot make observations for infinitesimally thin redshift bins.} Typically  a  top-hat window function is chosen \cite{Matthewson:2020rdt}.}
For the auto correlations, in order to keep numerical complications under control while introducing foreground filters, we only consider a Dirac window, which corresponds to $\Delta z\to 0$ and thus gives the theoretical angular spectra. 
The cross correlation \HI $\times\kappa$ is not affected by window functions \cite{Matthewson:2020rdt}, {since the CMB convergence field $\kappa$ is redshift independent. We checked this numerically.}

The remainder of the paper is structured as follows. We start by summarising foreground avoidance in Fourier space in \autoref{sec:ForgFilImple}, {where we also discuss the differences between the single-dish and interferometer modes of operation of HI IM surveys, including the effect of the beam in single-dish mode. A geometric interpretation of foreground filtering is developed in angular harmonic space, where all wide-angle effects are naturally included. Theoretical expressions for the generalised auto- and cross-angular power spectra with foreground filtering are derived in \autoref{sec:AngPowSpec}. We present numerical results in \autoref{sec:NumeResNDiscuss}. We show that non-diagonal correlations are introduced due to foreground filtering, but that these correlations are  small compared to the diagonal ones. We further show that the signal loss in  cross power is in general larger than in auto power. We conclude in \autoref{sec:Conclusion}.}

We assume a fiducial $\Lambda$CDM cosmology given by \textit{Planck} 2018 best-fit parameters \cite{Planck:2018vyg}: $A_\mathrm{s}=2.101\times 10^{-9}$, $n_\mathrm{s}=0.965$, $h=0.676$,  $\Omega_\mathrm{{b0}}=0.049$, $\Omega_\mathrm{{c0}}=0.261$. 

\section{Foreground filters in Fourier space} \label{sec:ForgFilImple}

HI IM  surveys can operate in two modes, as follows. 
\begin{itemize}
\item Single-dish (SD) mode:  auto-correlation signals from single dishes are added. This allows for probing large angular scales, {but small angular scales are lost due to damping by the telescope beam}.
\item Interferometer (IF) mode: cross-correlated signals from the array elements are combined, allowing for high resolution on small angular scales, but with the loss of large angular modes {due to a cut-off imposed by  the minimum baseline}.
\end{itemize}

{In the absence of instrumental noise, foregrounds and other systematics, the {\em theoretical} HI power spectra in SD and IF modes agree on intermediate scales which are not affected by the beam (SD) or minimum baseline (IF). However the two modes have (a)~different noise,  (b)~different effects from foregrounds, and (c)~different non-foreground systematics. This means that in practice the amplitude of the observed power spectra will differ between the two survey modes, even on intermediate scales where the theoretical power spectra agree.}

\subsection{{Survey properties}}

{We use nominal examples of HI IM surveys in order to illustrate our results. We consider an SD-mode survey similar to that planned for SKA-Mid \cite{SKA:2018ckk}, and an IF-mode survey similar to that planned for HIRAX \cite{Crichton:2021hlc,Newburgh:2016mwi}. {Hereafter, we refer to these as SKA-like and HIRAX-like surveys respectively.}
Since we are not producing forecasts, we do not need detailed specifications for these nominal surveys. We require the redshift range, which we assume to be $1\leq z\leq 3$ (this is realistic for SKA-Mid but extends the planned range of HIRAX). We also need the field of view
for dishes with diameter $D_\mathrm{d}$: \begin{equation}
\theta_\mathrm{b} = \frac{(1+z)\lambda_{21}}{D_\mathrm{d}}\,,
\quad\mbox{where}~~D_\mathrm{d}=15\,\mbox{m (SD)},~~6\,\mbox{m (IF)},
\label{eq:BeamSize}
\end{equation}
where the dish diameters correspond to those planned for SKA-Mid and HIRAX.}

{Estimates of the minimum and maximum angular scales accessible to the SKA-Mid and HIRAX-like surveys are determined in \cite{Durrer:2020orn} (see their Fig. 6), using the validity of linear perturbations, the effect of the field of view and the sky area. Based on their results, we assume the following multipole limits for our nominal surveys:
\begin{equation}
    \begin{array}{rll}
\mathrm{SD}: & \quad\ell_{\min }(z)=2, & ~~\ell_{\max }(z)
={450}/{(1+z)}, \\
\mathrm{IF}: &\quad \ell_{\min }(z)=150/(1+z), &~~ \ell_{\max }(z)
= 500\, .
\end{array}\label{eq:AngularScalesModes}
\end{equation}
}

{In the SD case, the maximum $\ell$ is imposed by the telescope beam. In fact, the beam damps the signal before the limit is reached, and we need to include this damping  in the  HI power \cite{Fonseca:2019qek}:
\begin{align}
C_\ell^{\mathrm{HI\times HI}}(z_1,z_2) &~~\longrightarrow ~~
\beta_\ell(z_1)\,\beta_\ell(z_2)\, C_\ell^{\mathrm{HI\times HI}}(z_1,z_2), \label{eq:BeamAuto}\\
C_\ell^{\mathrm{HI\times \kappa}}(z) &~~\longrightarrow~~ \beta_\ell(z)\, C_\ell^{\mathrm{HI\times \kappa}}(z), \label{eq:BeamCross}
\end{align}
where the angular power spectra $C^{A\times B}_\ell$ are defined in \autoref{sec:AngPowSpec} and \autoref{sec:NumeResNDiscuss}, and the beam factor is
\begin{equation}
    \beta_\ell(z) = \exp\left\{ -\frac{\ell(\ell+1)}{16\ln 2}\big[1.22\,\theta_\mathrm{b}(z)\big]^2\right\}.
\end{equation}
For the IF survey the minimum in \eqref{eq:AngularScalesModes} is imposed by the minimum baseline. The maximum $\ell_{\max }^{\rm IF}=500$ is imposed by the range of validity of linear perturbations. We replaced the redshift-dependent value in \cite{Durrer:2020orn} by a constant that is consistent with the limit required to safely avoid nonlinearity in the CMB lensing signal \cite{Lewis:2006fu}.}

\subsection{Filter properties}

The radial foreground filter  is applicable for both SD and IF surveys. It accounts for the loss of long-wavelength radial modes which are similar to the nearly spectrally smooth foregrounds (see \autoref{fig:FGCond}), leading to the Fourier space filter
\begin{equation}
\big|k_\| \big| >  k_{\rm Fg}=0.01\,h\,{\rm Mpc}^{-1} \quad \mbox{(SD+IF modes)},\label{eq:firstFGCond}
\end{equation}
where we use
a value of $k_\mathrm{Fg}$ that is often chosen for next-generation surveys {(e.g. \cite{Bull:2014rha,Sailer:2021yzm,Castorina:2019zho,Karagiannis:2019jjx,Cunnington:2020wdu, Spinelli:2021emp})}. 
Here we follow the standard assumption that $k_{\rm Fg}$ is redshift-independent, although there is in fact a weak $z$-dependence \cite{CosmicVisions21cm:2018rfq}.

The second filter applies to IF mode only. A baseline with a given physical length $L$ probes different angular scales $\ell$ at various frequencies. This implies that intrinsically monochromatic emission from a foreground point source can contaminate the spectrally non-smooth signal. If it is assumed that the contamination  operates within $N_{\rm w}$ primary beams of each pointing, then this leads to the exclusion of the primary beam `wedge' region in $(k_\perp,k_\|)$ space.  The Fourier space filter  is estimated as \cite{Liu:2014bba,Pober:2014lva,Alonso:2017dgh,CosmicVisions21cm:2018rfq, Karagiannis:2019jjx} (see \autoref{fig:FGCond}):
\begin{equation}
\big|k_\| \big| >\tan \alpha(z)\, k_\perp\quad \mbox{with}\quad
\tan \alpha(z)=
\frac{r(z)H(z)\sin\big[0.61\,N_{\rm w}\,\theta_\mathrm{b}(z)]}{(1+z)}
\qquad \mbox{(IF mode)}\,.\label{eq:wedge}
\end{equation}
Here $r$ is the comoving radial distance, $H$ is the Hubble rate and {$\theta_\mathrm{b}$} is defined in \eqref{eq:BeamSize}. 
{We assume that $N_{\rm w}=1$. 
Note that the wedge condition is not strictly a foreground-avoidance condition -- in principle, the wedge can be removed with excellent calibration of baselines \cite{Ghosh:2017woo, CosmicVisions21cm:2018rfq}, but {this is currently not possible with existing technology and} we treat it as futuristic.}

{Although the foreground filters are expressed in terms of $k_\|= k_z$ and $k_\perp^2 = k_x^2 + k_y^2$, they have a nice geometrical interpretation if we think in terms of $k_x$, $k_y$ and $k_z$ axes. 
\begin{enumerate}
\item SD Mode: Only \eqref{eq:firstFGCond} applies, and consequently  the region between the planes $k_z=\pm k_\mathrm{Fg}$  is discarded.
\item IF Mode: There is 
$\phi$ symmetry and thus the wedge line (blue line in \autoref{fig:FGCond}) upon rotation about the $k_{\|}\equiv k_z$ axis generates  a cone whose apex angle is $\pi/2-\alpha$. The allowed region \eqref{eq:wedge} is the interior of this cone -- and its reflection whose axis is the negative $k_z$ axis.
Combining with the radial filter \eqref{eq:firstFGCond}, 
the cones are truncated by the planes $k_z=\pm k_{\rm Fg}$.
\end{enumerate}}

\section{Angular power spectra \label{sec:AngPowSpec}}

{The foreground filters described above have been applied in the literature in Fourier space,  for a fixed line-of-sight direction, i.e., using a flat-sky approximation. In this section, we apply the filters in angular harmonic space, which is perfectly adapted to analyse fields on the sky, including all wide-angle correlations. We therefore do not assume a fixed line of sight, i.e.,  we do not impose a flat-sky approximation and thus our results encompass wide-angle correlations.}

For a field $X(z,\bm{n})$ on the sky, {where $\bm n$ is the line-of-sight direction to the field point,} the spherical harmonic expansion is  \begin{equation}
    X(z,\bm{n}) = \sum_{\ell\ge 0}\,\sum_{m=-\ell}^{\ell}a^X_{\ell m}(z)\,Y_{\ell m}(\bm{n}) \quad {\mbox{where}\quad
  a^X_{\ell m}(z)=\int {\rm d}\Omega_{\bm n}\,Y^*_{\ell m}(\bm n)\,X(z,\bm{n})} \,.
\end{equation}
Here $Y_{\ell m}(\bm{n})$ are spherical harmonics and $a^X_{\ell m}(z)$ are the spherical harmonic coefficients. 
{The field $X$ can be related to the primordial curvature perturbation $\zeta_{\bm{k}}$ by an angular transfer function in Fourier space, $\Delta^X_\ell(z,k)$, 
using the Rayleigh expansion}
\begin{eqnarray}
{{\rm e}^{{\rm i}\,\bm k \cdot \bm r} = 4\pi\sum_{\ell,m} {\rm i}^\ell\,j_\ell(kr)\, Y_{\ell m}\big(\bm n\big) \,Y^*_{\ell m}\big(\hat{\bm k}\big)
\quad\mbox{where}\quad \bm r=r\, \bm n\,.}
\end{eqnarray}
{Here $r$ is the line of sight distance and $j_\ell$ are spherical Bessel functions. Then it follows, as shown in detail in \cite{Bonvin:2011bg,Challinor:2011bk,DiDio:2013bqa,Fonseca:2018hsu}, that}
\begin{equation}
a^X_{\ell m}(z)=4\pi\, {\rm i}^\ell \int_{\mathcal{D}_X} \frac{\mathrm{d}^3\bm k}{(2\pi)^3}\, Y^*_{\ell m}(\hat{\bm k})\,\Delta^X_\ell(k,z)\,\zeta_{\bm{k}}\,.\label{eq:HarmCoeff}
\end{equation}
{For example, if $X$ is the HI brightness temperature contrast, then the dominant terms in $\Delta^{\HI}_\ell$ are given by matter density contrast  and redshift space distortion terms:
\begin{align}
\Delta^{\HI}_\ell(k,z)= b_{\HI}(z)\, j_\ell\big(kr(z) \big)\, {\cal T}_\delta(k,z) +  \frac{(1+z)}{H(z)} \,k\, j_\ell''\big(kr(z) \big)\, {\cal T}_{\bm v}(k,z)\,.
\end{align} 
Here $b_{\HI}$ is the HI bias, $\delta$ is the matter density contrast, $\bm v$ is the peculiar velocity, and ${\cal T}$ are the standard transfer functions in Fourier space:} 
\begin{equation}
{\delta_{\bm k}(z)=  {\cal T}_\delta(k,z)\,\zeta_{\bm k}\,,\quad  
  v_{\bm k}(z)=  {\cal T}_{\bm v}(k,z)\,\zeta_{\bm k}\,.}
\end{equation}

In \eqref{eq:HarmCoeff}, $\mathcal{D}_X(z)$ is the region of integration  in Fourier space. Thus we can study the effect of  foreground filtering by restricting the region of integration for \eqref{eq:HarmCoeff} as in
\autoref{fig:FGCond} for SD and IF modes.  
For later computation of angular power spectra, it is useful to write these regions in an explicit form. 
\begin{enumerate}
\item The CMB lensing region $\mathcal{D}_\kappa$ is not affected by HI foregrounds filters.
\item The HI region $\mathcal{D}_\HI$ follows from \autoref{fig:FGCond}. For both survey modes, $0\le\phi\le 2\pi$ and {$k>k_\mathrm{Fg}|\sec\theta|$}, where $\theta$ is restricted by 
\begin{eqnarray}
{  0\leq \theta\le \pi} &\qquad& \mbox{(SD mode)}\,, \\ { 0\leq \theta < \pi/2-\alpha~~\mbox{or}~~ \pi/2+\alpha< \theta\le \pi} &\qquad& \mbox{(IF mode)} \,.
\end{eqnarray}
\end{enumerate} 
Since $\alpha$ is $z$ dependent, so is the region of integration $\mathcal{D}_\HI(z)$.

With this, we can now calculate the correlations amongst two fields  $X,Y$:
\begin{align}
& \big\langle a^X_{\ell m}(z)a^{Y*}_{\ell'm'}(z')\big\rangle = (4\pi)^2 {\rm i}^{\ell-\ell'}\int_{\mathcal{D}_X(z)}\! \frac{\mathrm{d}^3\bm k}{(2\pi)^3} \int_{\mathcal{D}_Y(z')}\!\frac{\mathrm{d}^3\bm k'}{(2\pi)^3}  Y^*_{\ell m}(\hat{\bm k})\, Y_{\ell'm'} (\hat{\bm k}')
\notag 
\\ &\qquad\qquad\qquad \qquad\qquad\qquad\qquad\qquad\qquad~~~\times
\Delta^X_\ell(k,z)\,\Delta^Y_\ell(k',z')\,\big\langle \zeta_{\bm{k}}\zeta_{\bm{k}'}^* \big\rangle \,. \label{eq:2PCFHarmCoeff}
\end{align}
We can simplify \eqref{eq:2PCFHarmCoeff}, using $\big\langle \zeta_{\bm{k}}\zeta_{\bm{k}'}^* \big\rangle=(2\pi)^3\delta^{(3)}(\bm{k}-\bm{k}'){P_\zeta}(k)$, where $P_\zeta(k)$ is the primordial power spectrum: 
\begin{equation}
\big\langle a^X_{\ell m}(z)\,a^{Y*}_{\ell' m'}(z') \big\rangle = \frac{2}{\pi}\,{\rm i}^{\ell-\ell'}\int\mathrm{d}^3\bm k\, Y^*_{\ell m}(\hat{\bm k}) \,Y_{\ell'm'}(\hat{\bm k}) \, \Delta^X_\ell(k,z)\,\Delta^Y_\ell(k,z')\,P_\zeta(k)\,.
\label{eq:2PCFHarmCoeffSimp}
\end{equation}
Here {the integration is performed over the region common to both $\mathcal{D}_X(z)$ and} {$\mathcal{D}_Y(z')$}. 
Notice that in the absence of foreground filtering, {the common region} is the whole 3D space and we can perform angular integration which leads to $\delta_{\ell\ell'}$ and $\delta_{mm'}$ -- as expected in the absence of foregrounds. But in the presence of foregrounds this is no longer true.

Furthermore, since $\kappa\times\kappa$ is unaffected by HI foregrounds, we only study the HI auto ($\HI\times\HI$) and cross ($\HI\times\kappa$) correlations.
In practice, {observations of the HI brightness temperature average over finite redshift bins, which are defined by window functions}. 
The harmonic coefficient   for any given window function $W(z,\tilde{z})$, centred at  redshift $z$, follows from \eqref{eq:HarmCoeff}:
\begin{equation}
a^\HI_{\ell m}(z) = 4\pi\, {\rm i}^\ell \! \int \!
\mathrm{d}\tilde{z}\, W(z,\tilde{z})\, \Delta^\HI_\ell(k,\tilde{z})  \int_{\mathcal{D}_\HI}\, \frac{\mathrm{d}^3\bm k}{(2\pi)^3}\, Y^*_{\ell m}(\hat{\bm k})\, \zeta_{\bm{k}}\,.
\label{eq:WindowAutoPower}
\end{equation}
However we find that significant computational complications arise with window functions when including foreground filters, and for our purposes here 
we show numerical results using only a Dirac window, {$W(z,\tilde z)=\delta^{\rm D}(z-\tilde z)$}.

\subsection{\HI\ auto power}

Integration over $\phi$ in \eqref{eq:2PCFHarmCoeffSimp} gives $\delta_{mm'}$ due to axial symmetry. {Then the HI correlations take the form
\begin{align}
\boxed{\big\langle a^\mathrm{\HI}_{\ell m}(z) \, a^{\mathrm{\HI}*}_{\ell' m'}(z') \big\rangle = f^{m,m}_{\ell,\ell'}\delta_{mm'}\int_{Z(z,z')}^{1}\mathrm{d}x\, P^m_\ell(x)\,P^m_{\ell'}(x)\,\mathcal{C}^{\HI,\HI}_{\ell\ell'}(z,z',x),}
\label{eq:2PCFAutoHarmCoeff}
\end{align}
where we defined
\begin{equation}
\boxed{\mathcal{C}^{XY}_{\ell\ell'}(z,z',x) =\frac{2}{\pi}
\int_{k_{\mathrm{Fg}}/x}^\infty\mathrm{d}k\, k^2 \, \Delta^{X}_\ell(k,z)\, \Delta^Y_{\ell'}(k,z') \,P_\zeta(k)\,.}
\end{equation}}
Here $P^m_\ell(x)$ are associated Legendre polynomials, the lower limit of the $x$ integral is 
\begin{align}
Z(z,z') &=
\begin{cases}
\pi/2 & \text{for SD,} \\ \max
{\big\{ \sin \alpha(z),\,\sin \alpha(z')\big\}}
& \text{for IF,}
\end{cases}
 \label{eq:XIntLimit} 
\end{align}
and
\begin{align}
f^{m,m'}_{\ell,\ell'} = \frac{{\rm i}^{\ell-\ell'}}{2}\Big[1+(-1)^{\ell+\ell'+m+m'}\Big]\left[{\frac{(2\ell+1)(2\ell'+1)(\ell-m)!(\ell'-m')!}{(\ell+m)!(\ell'+m')!}} \right]^{1/2}.
\label{eq:MultFact}
\end{align}\\

\subsection{HI$\times \kappa$ cross power}
For the cross correlation, as pointed out above, $\mathcal{D}_\kappa$ is the whole of  Fourier space. Thus \eqref{eq:2PCFHarmCoeffSimp} simplifies to
\begin{equation}
{\boxed{\big\langle a^\mathrm{\HI}_{\ell m}(z)a^{\kappa*}_{\ell' m'} \big\rangle =  f^{m,m}_{\ell,\ell'}\delta_{mm'}\int_{{\sin\alpha(z)}}^{1}\mathrm{d}x\, P^m_\ell(x) \,P^m_{\ell'}(x) 
\,\mathcal{C}^{\HI,\kappa}_{\ell\ell'}(z,x).
}}
\label{eq:2PCFCrossHarmCoeff}
\end{equation}
In the case of SD mode,  ${\alpha= 0}$.
\vspace*{0.2cm}


We highlight some of the common features shared by expressions \eqref{eq:2PCFAutoHarmCoeff} and \eqref{eq:2PCFCrossHarmCoeff}:
\begin{itemize}
\item There is an explicit $m$ dependence and both expressions are invariant under $m\to -m$. This is a consequence of $\phi$ symmetry in  Fourier space.
\item There are off-diagonal correlations, i.e., $\Delta\ell=\ell'-\ell\ne0$,  due to the loss of $\theta$ symmetry in  Fourier space (explicit in \autoref{fig:FGCond}).
\item From \eqref{eq:MultFact}, it is clear that  $f^{m,m}_{\ell,\ell'}\ne0$ only when  $\ell+\ell'$ is even. This implies that $\Delta\ell=\ell'-\ell$ is also even, ensuring that both expressions \eqref{eq:2PCFAutoHarmCoeff} and \eqref{eq:2PCFCrossHarmCoeff} are real.
\end{itemize}

Given an $m$ dependence and the presence of non-diagonal correlations in \eqref{eq:2PCFAutoHarmCoeff} and \eqref{eq:2PCFCrossHarmCoeff}, we can define the \textit{generalized} angular power spectra for both auto and cross correlations in the presence of foreground filters:
\begin{equation}
C^{XY}_{\ell\ell'}(z,z')=\sum_{m=-L}^{L} \, \frac{\big\langle a^X_{\ell m}(z)\, a^{Y*}_{\ell' m}(z')\big \rangle}{\sqrt{(2\ell+1)(2\ell'+1)}}\quad\text{where}\quad L=\min\{\ell,\ell'\}\,.
\label{eq:GenAngPow}
\end{equation}
In the absence of foregrounds, i.e., when $k_\mathrm{Fg}\to 0$ and ${\alpha\to 0}$, this expression takes the expected form,
\begin{equation}
C_{\ell}^{XY}(z,z')=\frac{2}{\pi}\int_{0}^{\infty}\mathrm{d}k \,k^2 \,\Delta^X_\ell(k,z) \, \Delta^Y_{\ell}(k,z') \,P_\zeta(k) \,.
\end{equation}
\begin{figure}
\centering
\includegraphics[width=14cm]{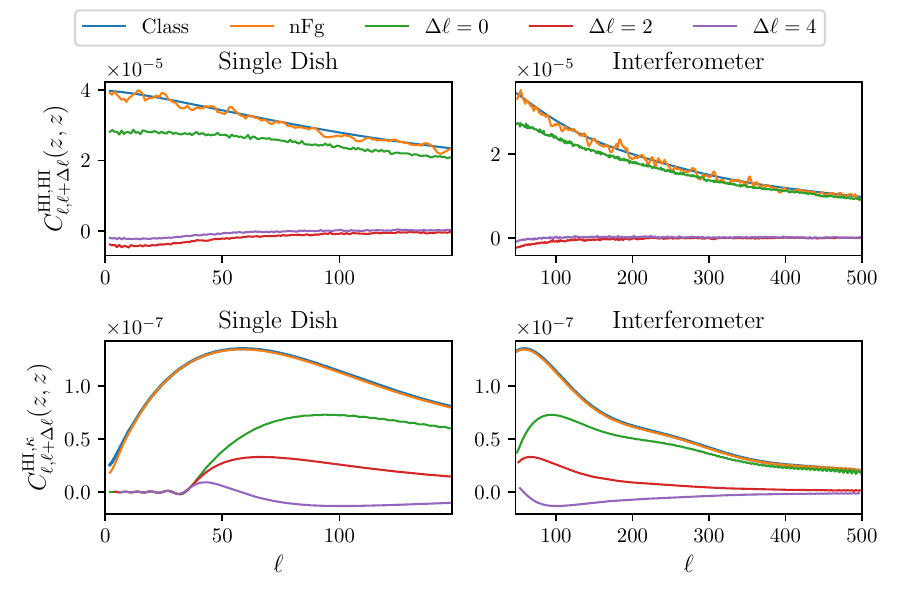}
\caption{Effect of $\Delta \ell$ on the auto power $\HI\times\HI$ (top) and cross power $\HI\times\kappa$ (bottom), for
single-dish (left) and interferometer (right) surveys, at {$z=2$. {We also make a comparison between our code (orange curves, `nFg') and the \texttt{Class} code (blue curves) code in the absence of foregrounds.} 
Our code agrees with \texttt{Class}, especially for the cross power; for auto power, numerical complications cause oscillations, even in the absence of foregrounds. Green, magenta and red curves show the effect of foreground filters for $\Delta \ell=0,2,4$ respectively.} (We use a Savgol filter \cite{doi:10.1021/ac60214a047} of order 4 and window size 15 for smoothing.)}
\label{fig:VaryDelEll}
\end{figure}

\section{Numerical results and discussion \label{sec:NumeResNDiscuss}}

{Before discussing our numerical results, we should also mention that we have benchmarked our code with \texttt{Class} as the results from both should match in the absence of foregrounds (see the orange and blue curves in \autoref{fig:VaryDelEll}). For the cross power, excellent agreement is found. In case of auto power, on account of numerical complications, there are fluctuations but the match is still good.}

\begin{figure*}
\centering
\includegraphics[width=16cm]{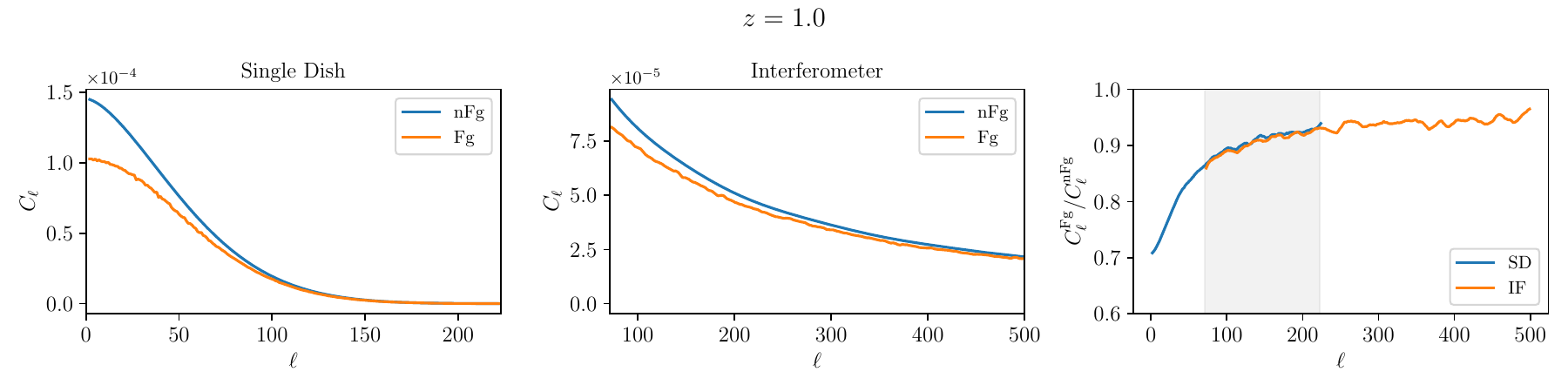}\\
\includegraphics[width=16cm]{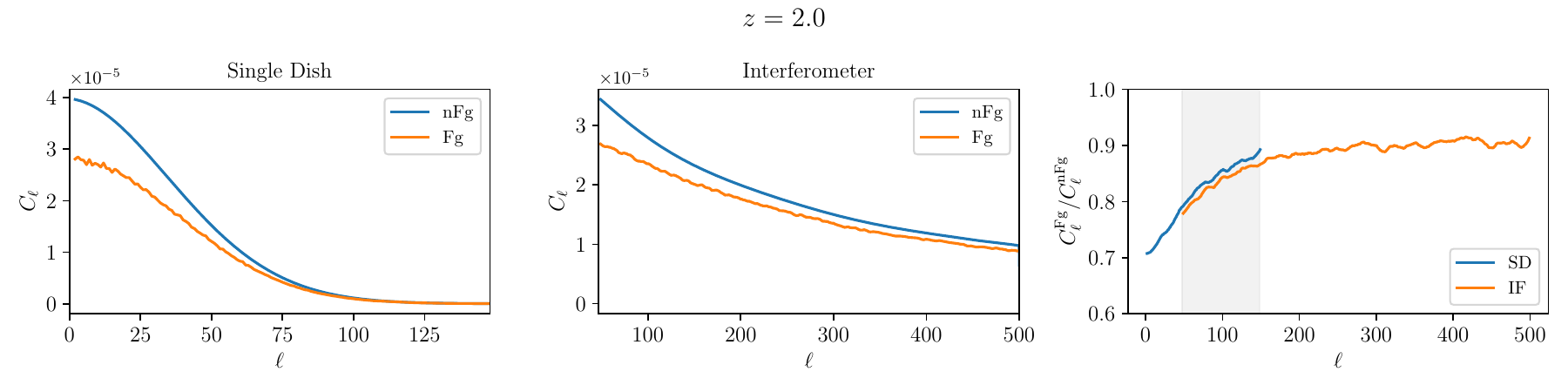}\\
\includegraphics[width=16cm]{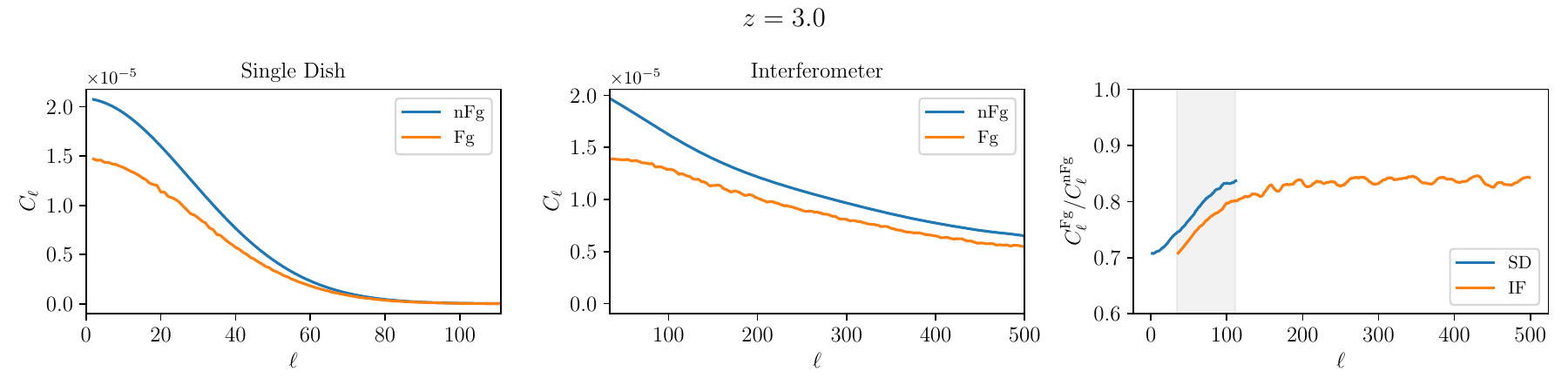}
\caption{Foreground filtering effects on the {auto} HI power in SD and IF modes at {$z= 1, 2, 3$}. {The range of $\ell$ values are chosen following \eqref{eq:AngularScalesModes}. For each redshift, the first panel compares the SD power spectra, including the telescope beam as in \eqref{eq:BeamAuto}. It can be seen that the beam damps the signal with increasing strength as $z$ increases.  The middle panel does the same for IF {mode}. The third panel compares the ratio $C^\mathrm{Fg}_\ell/ C^\mathrm{nFg}_\ell$ for the two modes of survey. It is apparent that in the overlap range (shaded region) the power loss in SD mode is smaller than or equal to that  in IF mode.}  
\label{fig:FGEffectAuto}}
\end{figure*}

\subsection{Comparison of diagonal and off-diagonal terms}
{In \autoref{fig:VaryDelEll}, we study the effect of $\Delta\ell$ on both auto and cross power spectra, as given by \eqref{eq:GenAngPow}. The angular scales, in  SD and IF modes and for auto and cross spectra, are chosen as per the discussion in \S \ref{sec:ForgFilImple}. Beam effects are not included in this plot since our objective is only to compare relative magnitudes of power spectra for various $\Delta\ell$ values.}

{In the top and bottom panels, we respectively show the auto power $C_{\ell,\ell+\Delta\ell}^{\HI,\HI}(z,z)$ and cross  power $C_{\ell,\ell+\Delta\ell}^{\HI,\kappa}(z)$ at $z=2$, for different $\Delta\ell$. The left and  right panels are respectively for SD and IF modes. For the auto power $\HI\times\HI$, it is evident that the non-diagonal correlations $\ell\ne\ell'$  are neglible  compared to the diagonal ones: we can safely neglect the non-diagonal correlations and take $\ell =\ell'$ in \eqref{eq:GenAngPow} for $\HI\times\HI$. In the cross-power case, $\HI\times\kappa$, this also holds for the $\Delta\ell=4$ contributions, but the $\Delta\ell=2$ contributions are significant for a small $\ell$ interval on the largest available scales. For an accurate treatment, it would be necessary to include these off-diagonal contributions, but for our purposes we will neglect them.}

\begin{figure*}
\centering
\includegraphics[width=16cm]{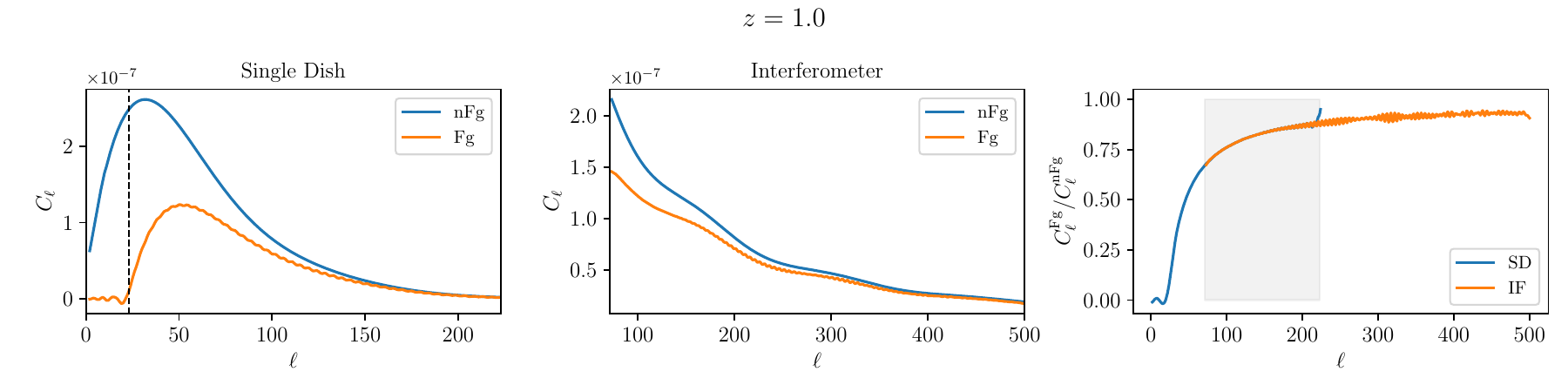}\\
\includegraphics[width=16cm]{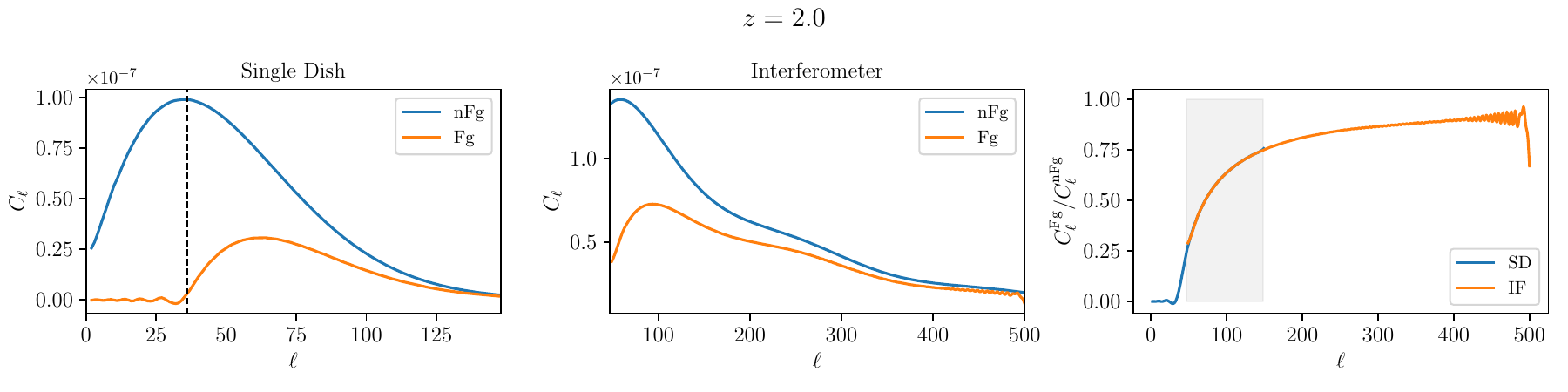}\\
\includegraphics[width=16cm]{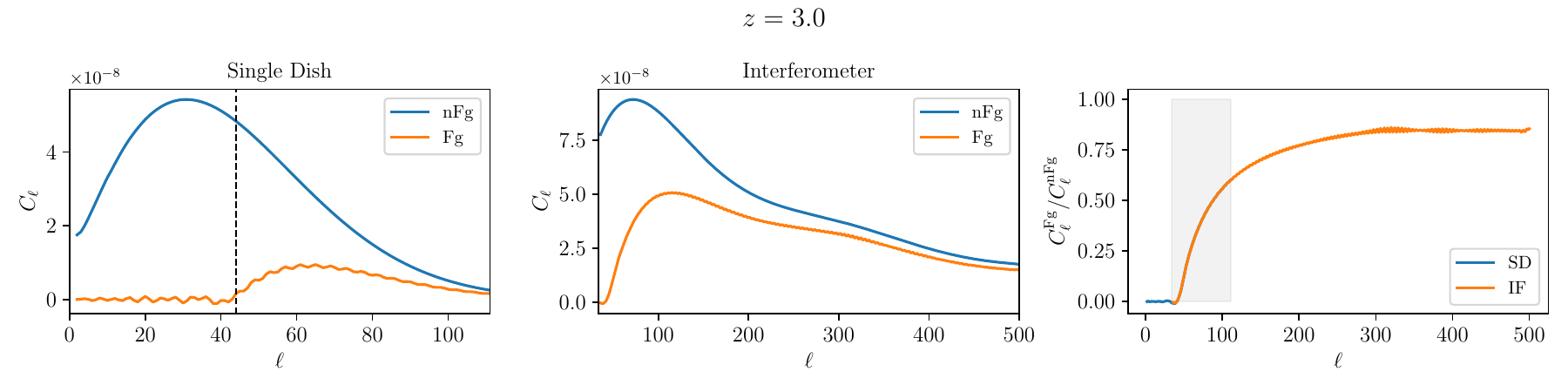}
\caption{As in \autoref{fig:FGEffectAuto} but for the cross-power $\HI\times\kappa$, {including the HI telescope beam as in \eqref{eq:BeamCross}.} 
 {In the first panels for each $z$, vertical dashed lines show the locations of $\ell_0$, estimated using \eqref{eq:EllMinLoca}. It is evident that the estimate of $\ell_0$ is quite accurate. The third column of panels shows that the power loss is the same for both SD and IF modes in the overlap range.
 (On account of large oscillations at high $\ell$  we use a Savgol filter of larger window size 35 and order 4.)}  
\label{fig:FGEffectCross}}
\end{figure*}
\begin{figure}[!h]
\centering

\includegraphics[height=6.2cm]{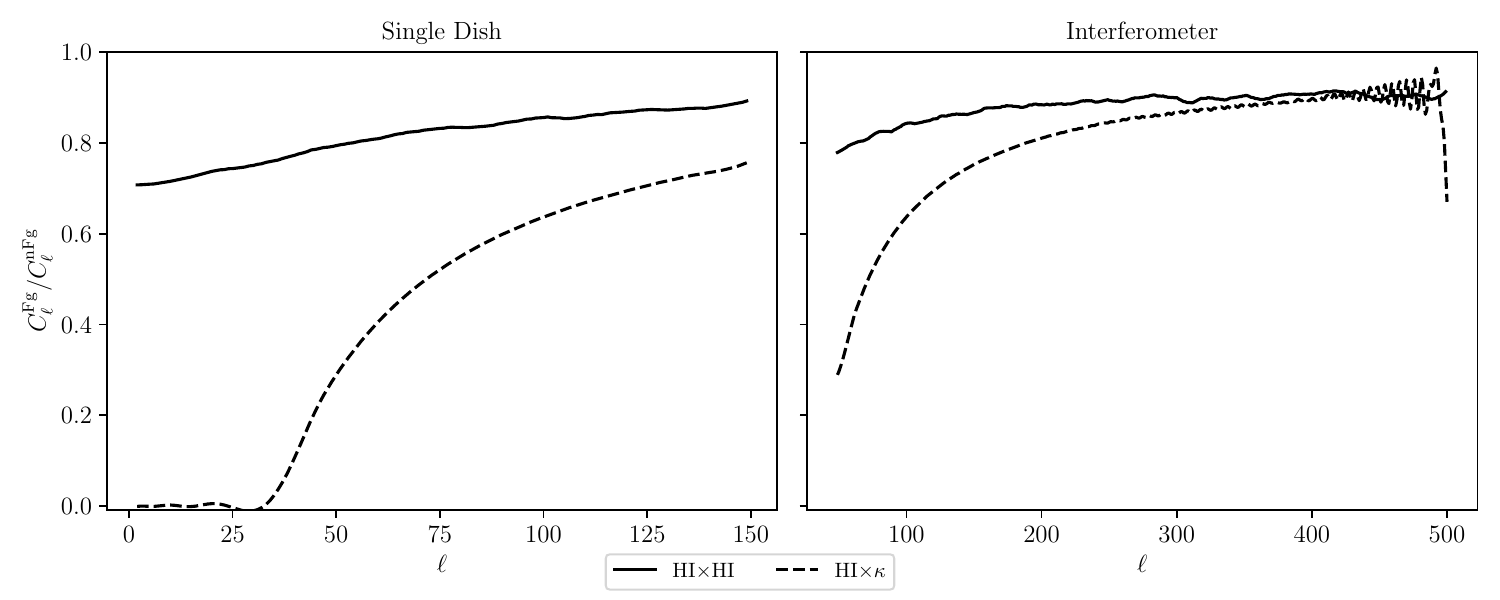}
\caption{Comparison of HI$\times$HI and HI$\times\kappa$ power losses due to foregrounds at {$z=2.0$}. {For both  auto and cross power spectra, and in both modes of survey,  $C_\ell^\mathrm{Fg}/C_\ell^\mathrm{nFg}$ increases with $\ell$, so that the power loss decreases as $\ell$ increases. 
The power loss for HI$\times\kappa$ is always $\geq$  that for HI$\times$HI.
}} 
\label{fig:AutCrosPowLossCom}
\end{figure}

\subsection{Power loss due to foreground filters}

We further investigate
both auto and cross power spectra in the presence of foregrounds, {including the effect of the telescope beam in SD mode and of the minimum baseline in IF mode. Our findings are:}

\begin{enumerate}
\item {For auto spectra, \autoref{fig:FGEffectAuto} shows that the power loss in IF mode is similar to that of SD mode in the overlap region, i.e., the region of common $\ell$ values. Although the losses in both spectra increase as $z$ increases, the power loss becomes more prominent for IF mode as compared to SD. This can be  understood from  \eqref{eq:XIntLimit}, since increasing $z$ leads to a larger value of $\sin\alpha$, which in turn reduces the integration range in \eqref{eq:2PCFAutoHarmCoeff}}. 
\item {In the cross-power case, HI$\,\times\,\kappa$, \autoref{fig:FGEffectCross} shows the same trend of power reduction as a function of $z$. But in contrast to the auto spectra, the power reduction in the overlap region for both SD and IF modes is the same.}
\item Furthermore the HI$\,\times\,\kappa$ signal vanishes for {$\ell<\ell_{0}$. Since $\ell_0<\ell_{\min}^{\rm IF}$, this feature is not seen in IF mode, but only in SD mode.}
 Further, since $0\le x\le1$ [see \eqref{eq:2PCFCrossHarmCoeff}], we have $k>k_\mathrm{Fg}/x > k_\mathrm{Fg}$. Then a rough estimate of {$\ell_{0}$} is 
 \begin{equation}
{\ell_{0}(z)}\sim k_\mathrm{Fg}r(z)\sim \big(23, 36, 44 \big) ~~\mbox{for}~~   z=  \big(1,2,3 \big).  \label{eq:EllMinLoca}
 \end{equation}
  These estimates match reasonably well with the values evident from  \autoref{fig:FGEffectCross} {(left-most panels)}. 

\item The HI$\,\times\,\kappa$  power is more severely affected by foregrounds than the  HI$\,\times\,$HI power {in both survey modes},  
for all redshifts used in our analysis. In \autoref{fig:AutCrosPowLossCom}, we make a comparison of the power loss for SD and IF modes {at $z=2$}. {For both modes of survey, and for auto- and cross-power, the power loss decreases for larger $\ell$ values.} 
When $\ell\sim 500$, the power losses in IF mode for the auto and cross spectra 
 approach the same value.
\end{enumerate}

\section{Conclusion and outlook \label{sec:Conclusion}}

{HI intensity mapping offers a powerful new probe of cosmology. However, the exploitation of this probe requires the removal of very large foreground contamination. In this paper, we present a more modest treatment, in which filters are applied that aim to avoid the regions of serious foreground contamination. This can give an indication of what is possible, but it is clearly not a substitute for a comprehensive analysis and data pipeline that incorporate foreground cleaning.}

We develop a new geometrical interpretation of foreground filtering for HI intensity mapping in harmonic space. {Our analysis does not depend on a flat-sky approximation (with its fixed line of sight) and therefore incorporates all wide-angle correlations. The main analytical results are given in \eqref{eq:2PCFAutoHarmCoeff}--\eqref{eq:2PCFCrossHarmCoeff}, showing the generalised power spectra $C_{\ell,\ell'}^{\HI,\HI}$ and $C_{\ell,\ell'}^{\HI,\kappa}$.}

In order to keep computational complications under control, we only present numerical results for the Dirac window choice {in \eqref{eq:WindowAutoPower}}. The cross power $\HI\times\kappa$ is independent of window functions. 

As an application, we  study the foreground-filtered $\HI\times\HI$ and  $\HI\times\kappa$ spectra, where $\kappa$ is the CMB convergence field. {We can summarise our findings as follows.

\begin{itemize}
\item Foreground filtering in harmonic space leads to a  loss of symmetry and consequently to anisotropic angular power spectra in which there are off-diagonal correlations, {$C_{\ell,\ell'}\neq 0$ for $\Delta\ell\equiv\ell'-\ell=2,4,\cdots$. For the auto power $\HI\times\HI$, these off-diagonal correlations are negligible, as illustrated in \autoref{fig:VaryDelEll}.  In the cross power, $\HI\times\kappa$, the $\Delta\ell=2$ contribution is not negligible for a small interval of the largest available scales (lowest $\ell$). A more accurate treatment of the cross power should include this contribution.}
\item 
{When neglecting the $\ell'\neq\ell$ contributions and  {instrumental} noise, our numerical results are well illustrated by \autoref{fig:FGEffectAuto}--\autoref{fig:AutCrosPowLossCom}.}
\item The power losses due to foreground filtering  for both auto- and cross-power, increase with redshift $z$, as expected.
\item For a given  $z$, harmonic multipole $\ell$ and observing mode (single-dish or interferometer), the power loss in auto-correlations HI$\times$HI is less than {or equal to that in} cross-correlations \HI $\times\kappa$.
\item {The power loss for the interferometer mode of operation is equal to, or slightly greater than, the loss for  single-dish mode, on the multipole interval where both modes have signal. The approximate equality may seem surprising, since
the foreground wedge avoidance affects only interferometer mode. However, this wedge loss only applies to higher multipoles, where the single-dish telescope beam has already effectively wiped out the signal.
(Note also that we neglect the thermal noise, which is different for single-dish and interferometer modes.)}
\item The $\HI\times\kappa$ power spectrum vanishes {in single-dish mode for $\ell<\ell_0$, where $\ell_0$ is estimated in \eqref{eq:EllMinLoca}. This feature is not seen in IF mode since $\ell_0<\ell_{\min}^{\rm IF}$.}
\item
{The loss of power in $\HI\times\kappa$, due to   foreground avoidance, is less severe than in a Fourier analysis which uses the flat-sky approximation, e.g. \cite{Moodley:2023lmu} (see their Fig. 1). The flat-sky approximation neglects wide-angle correlations of HI intensity and CMB convergence at each redshift -- whereas our angular cross-power spectrum includes these correlations, which reduces the loss of signal. We intend to further investigate this in a follow-up work.}
\end{itemize}
}

Finally, our techniques are general and may be useful to derive wide-angle analytical expressions for foreground-filtered angular \HI\ auto- and cross-bispectra {(see e.g. \cite{DiDio:2018unb,  Durrer:2020orn, Kothari:2021pbp}  
and references therein)}.

\newpage

\noindent{\bf Acknowledgments}\\
We thank Mario Ballardini, William Matthewson, Adrian Liu, Prabhakar Tiwari, Sandeep Kushwah and Richa Garg for useful comments and discussions, {and the anonymous reviewer for comments that helped to improve the paper}.  
The authors acknowledge  support from the South African Radio Astronomy Observatory and  National Research Foundation (Grant No. 75415).  We used the HPC facilities at IIT Mandi (Grant No IITM/SG/DIS-ROS-SPA/111) and  at the South African Centre for High Performance Computing (under the project {\em Cosmology with Radio Telescopes}, ASTRO-0945), the code \texttt{Class}, and the python packages \texttt{scipy}, \texttt{astropy}  and \texttt{Camb}.

\bibliographystyle{jhep}
\bibliography{reference}

\end{document}